**Side-gate leakage and field emission in all-graphene field effect transistors on SiO$_2$/Si substrate**

A. Di Bartolomeo[1,2,*], F. Giubileo[2], L. Iemmo[1,2], F. Romeo[1,2], S. Russo[3], S. Unal[3], M. Passacantando[4], V. Grossi[4], A. M. Cucolo[1,2]

[1] Physics Department "E.R. Caianiello", University of Salerno, via Giovanni Paolo II, 132, 84084, Fisciano, Salerno, Italy

[2] CNR-SPIN, via Giovanni Paolo II, 132, 84084, Fisciano, Salerno, Italy

[3] Physics Department, Exeter University, Exeter, EX4 4QL, UK

[4] Department of Physical and Chemical Sciences, University of L'Aquila, Via Vetoio, Coppito, L'Aquila, Italy

[*] Corresponding author. E-mail: dibant@sa.infn.it

**Abstract**

We fabricate planar all-graphene field-effect transistors with self-aligned side-gates at 100 nm from the main graphene conductive channel, using a single lithographic step. We demonstrate side-gating below 1V with conductance modulation of 35% and transconductance up to 0.5 mS/mm at 10 mV drain bias. We measure the planar leakage along the SiO$_2$/vacuum gate dielectric over a wide voltage range, reporting rapidly growing current above 15 V. We unveil the microscopic mechanisms driving the leakage, as Frenkel-Poole transport through SiO$_2$ up to the activation of Fowler-Nordheim tunneling in vacuum, which becomes dominant at high voltages. We report a field-emission current density as high as 1µA/µm between graphene flakes. These findings are essential for the miniaturization of atomically thin devices.

---

# 1. Introduction

Since the discovery of graphene [1], the field effect modulation of the relativistic charge carriers in this atomically thin material has largely been investigated in capacitive structures whereby a highly doped substrate coated by an oxide dielectric was used as back-gate to apply a vertical electric field. Extra control was then added by including a top-gate, at the cost of process complexity and increased risk of device failure due to the top dielectric deposition [2-5]. The use of side-gate(s) to limit interaction with dielectrics and avoid mobility degradation emerged soon after with side-gates formed by graphene [6-10] or metal leads [11]. All-graphene devices, where graphene is used both as channel and side-gate, offer the further advantage of the fabrication of self-aligned structures in a single lithographic step, with optimized gate to source/drain overlapping.

To date, research on side-gated devices has been dealing with the effect of the gate on the electrical transport properties of the graphene channel in terms of the modulation of conductance [6-7,11], penetration of the transversal field in the channel [6], transconductance [8]. Different geometries with single and dual side-gate and various critical dimensions have been considered [9], and some progress has been made towards a process suitable for integration in the existing Si technology [10].

The potentially high gate leakage caused by current flowing from gate to channel through the dielectric/air or dielectric/vacuum interface is one of the main weaknesses of graphene side-gated transistors [7]. Besides, the electron transport between horizontal graphene flakes at a nanometric distance and on a dielectric surface is by itself an interesting fundamental problem. Yet, it has received almost no attention. Although electron emission from the edge of a graphene flake has been largely investigated in connection to the quest for graphene field emission devices [12-18], the current between two graphene flakes separated by a nanogap has been rarely investigated. Wang et al. [19] patterned graphene sheets with crystallographically matching edges by divulsion, separated by a few hundred-nanometers distance on $SiO_2$/Si and used them to measure the current-voltage (I-V) characteristics in a high-vacuum chamber. They found that I-V curves are governed by the space-charge-limited flow of current at low biases and by Fowler–Nordheim [20] tunneling in high voltage regime. A recent study on field emission between suspended graphene flakes also reported Fowler-Nordheim field emission as a dominant microscopic mechanism and a current density as high as 10 nA/µm at modest voltages of tens of volts [21]. In particular, it was found that the emission is stable in time and repeatable over large numbers of voltage cycles, and that the emission current follows a power law dependence on pressure, a feature that they suggest to exploit for sensing purposes.

In this paper, we study the side-gating effect and the gate leakage in all-graphene devices. We fabricate side-gated all-graphene field-effect transistors by patterning exfoliated flakes on a $SiO_2$/Si substrate. Channel and gate are formed by graphene with parallel edges, separated by 100 nm gap. Such configuration helps to build a uniform electrical field along the channel direction, which is suitable to study the planar gate-to-channel leakage mechanisms and the electron field emission from individual graphene flakes. We report rapidly growing leakage for an electric field higher than 150 V/µm and we clarify that the current is due to Frenkel-Poole transport through $SiO_2$ until Fowler-Nordheim emission in vacuum between graphene flakes takes over and dominates with a current reaching a value as high as 1µA/µm at 100V.

Our study offers opportunities for both fundamental and applied research in vacuum nanoelectronics.

## 2. Experimental setup

Graphene devices were fabricated by standard electron-beam lithography (EBL) on heavily p-doped Si substrates (resistivity is 0.001-0.005 Ω·cm) capped with a 290 nm thermally-grown $SiO_2$ layer. Single and bilayer graphene flakes were exfoliated from HOPG and identified under optical microscope using contrast analysis under green light [22-23]. The selected graphene flakes (Figure 1(a)) were patterned in a single lithographic step to define channel and gate. Unwanted graphene was removed by $O_2$ plasma etch (Figure 1 (b) and (c)). Metal contacts (20 nm Cr/90 nm Au) were defined in a successive EBL step by metal evaporation and lift-off (Figure 1(d)). To achieve low contact resistance, metal was evaporated at a low rate (0.2-0.4 Å/s for the Cr adhesion layer and at 0.4-0.9 Å/s for the capping Au), at a pressure of $4-8\times10^{-6}$ Torr.

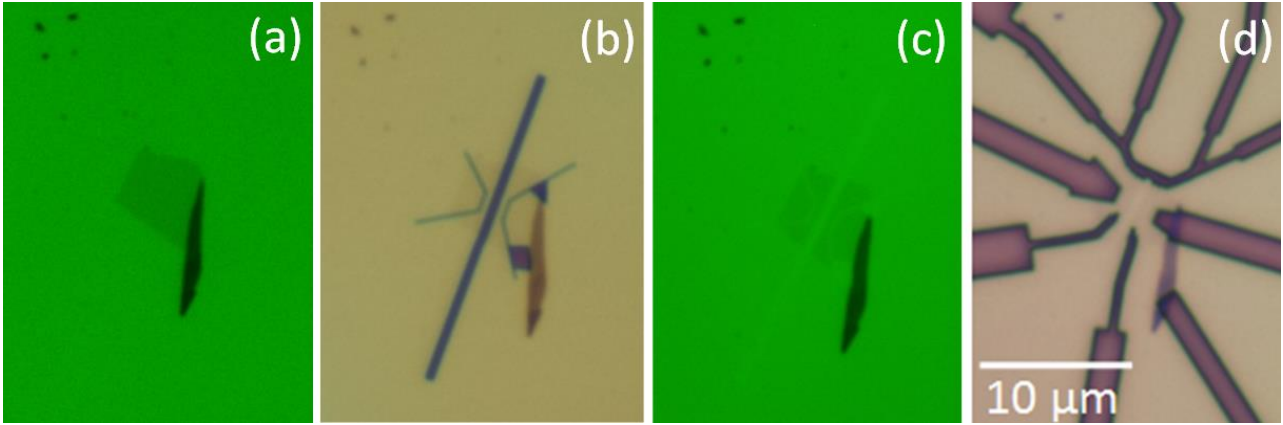

FIG. 1. Optical image (under green light and magnification 50×) of an exfoliated graphene bilayer as deposited (a), covered by developed PMMA after e-beam exposure to define gate and source/drain regions (b), as patterned after $O_2$ plasma exposure and PMMA removal (c), covered by developed PMMA after e-beam exposure for contact definition.

The layout/schematic and the scanning electron microscope (SEM) image of a typical device are shown in Fig. 2(a) and 2(b). The device consists of two side-gated transistors on a Si substrate (back gate), connected in series by a metal line. The two side gates are controlled separately and can be biased to make the two transistors p and n-type, respectively, thus enabling a CMOS all-graphene device.

The side-gate architecture offers easy control of the transistor dimensions, as the gate-to-channel distance or the channel width, which are only limited by the resolution of the lithography/etching process. In our design, we enlarged the source/drain regions to reduce the external resistance. Furthermore, we shaped the gate to maximize its overlapping and prevent the formation of unwanted high resistive paths, often included in top-gated devices. To achieve devices for low-power applications working with biases below 1V, we chose a gate-to-channel distance $d$=100 nm and a channel width $w$=500 nm.

Electrical measurements were performed in a Janis probe station at a pressure of ~1 Torr. To achieve $SiO_2$/vacuum interface as gate dielectric, for the gate leakage and field emission measurements, we placed the sample in a nanoprobe-equipped Zeiss SEM chamber at ~$10^{-6}$ Torr. In both cases, we used a Keithley 4200 SCS as source-meter unit.

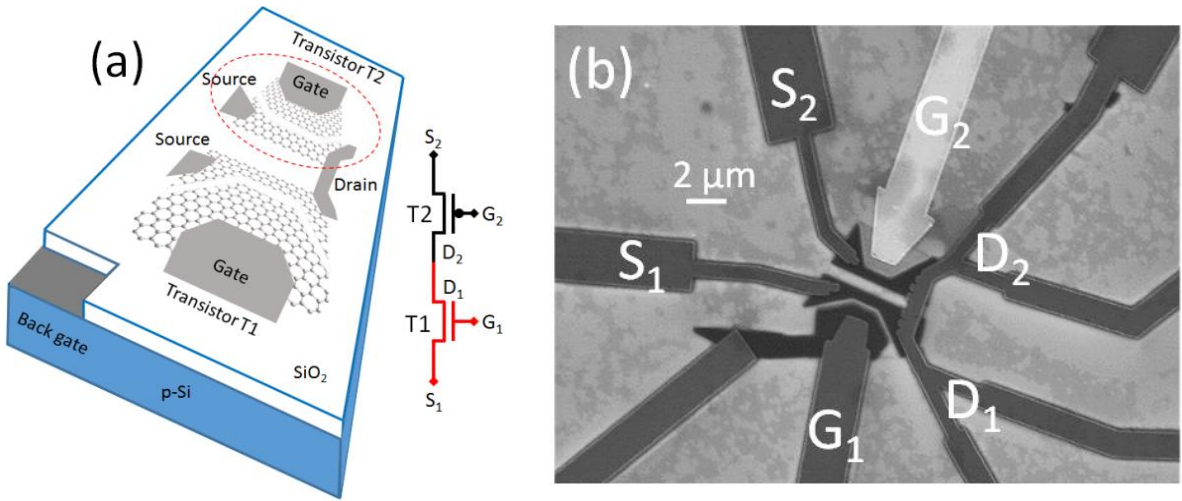

FIG. 2. (a) Layout and schematic of two side-gated graphene transistors (T1 and T2) connected in series. The transistors share the back-gate and are endowed with independent side-gates. (b) SEM image of the device in (a) showing electrically connected metal leads (darker lines; the gate $G_2$ of T2 is floating, thus appears lighter in the SEM image).

## 3. Results and discussion

Figure 3 shows the effect of the side-gate, and of the back-gate for comparison, on the channel conductance of the single transistor. The drain current-voltage relation ($I_D$-$V_D$ curve, Fig. 3(a)), measured at different side-gate biases $V_G$ and for grounded source and floating back-gate, shows a linear behavior. The minimum source-to-drain resistance, which includes the channel and the two contact resistances, is ~50 kΩ and is dominated by the channel. Measurements on test structures fabricated simultaneously and on the same chip of the device of Fig. 2 confirmed a contact resistance of the order of 1 kΩ/μm, consistent with previous experiments [24-28]. The modulation of the graphene conductance G by the side-gate is shown by the transfer characteristic (G-$V_G$ curve) displayed in Fig. 3(b), which shows a 35% variation in an interval of ~0.5 V. As common for air and PMMA exposed graphene transistors [28], the neutrality (Dirac) point is located at positive $V_G$ (~0.35 V), which indicates a p-channel device. The inset of the Fig. 3(b) shows the conductance behavior in a full loop of $V_G$ values, and shows a significant hysteresis between the forward and the reverse sweep. We characterize the hysteresis by the ratio of the forward-reverse voltage shift $\delta V_G$ to the width $\Delta V_G$ of the V-shaped curve, at a given G, and find $\delta V_G / \Delta V_G |_{G=15\mu S} \approx 20\%$. For comparison, Fig. 3(c) shows the transfer characteristic generated by the back-gate, with floating side-gate, and we observe a similar hysteretic behaviour with higher $\delta V_G / \Delta V_G |_{G=15\mu S} \approx 39\%$ (see inset of Fig 3(c)). The hysteresis is attributed to charge transferred and stored in charge traps present in the gate dielectric as well as induced by polymer residues in the processing or by unwanted contamination – e.g. adsorbates or moisture [29-32]. Residues, adsorbates or moisture may result particularly dangerous for side-gate devices, especially if localized at the graphene edges or in the channel-gate spacing, since they can form dipoles, which disturb the local electrical field and deteriorate the gating effect, or contribute to the gate leakage current at higher electric fields. We performed measurements in vacuum and kept the side-gate bias low to prevent leakage, which was always below the floor noise of the experimental setup.

Fig. 3(d) shows the transconductance, $G_m = dI_D / dV_G |_{V_D=9mV}$, normalized by channel width [8], for side- and back-gate. The transconductance is a useful parameter in the saturation regime, when a FET is used as an amplifier [39]. Here, saturation is not achieved; nonetheless, we use the transconductance to compare the side- and back-gate ability to convert a voltage to current. Fig. 3(d) shows that in the narrow interval |$V_G$|<1V the side-gate efficiency is 5 to 10 times higher than the back-gate on a 50 times larger voltage interval.

The back-gate sweep (Fig. 3(c)) confirms the p-type behavior of graphene with a Dirac point at $V_G \approx$ 10 V. The wider sweeping range evidences another feature, which often appears in back-gated transistors, i.e. a second Dirac point. This feature has been explained in terms of graphene doping by the contacts and Fermi level de-pinning [32-38]. A careful comparison of the shape of the curves in the insets of Fig. 3(b) and 3(c), over the same current range, seems to exclude the appearance of a second dip in the side-gate transfer characteristics, and, in fact, we did not find any evidence of it over a sweep up to 3V.

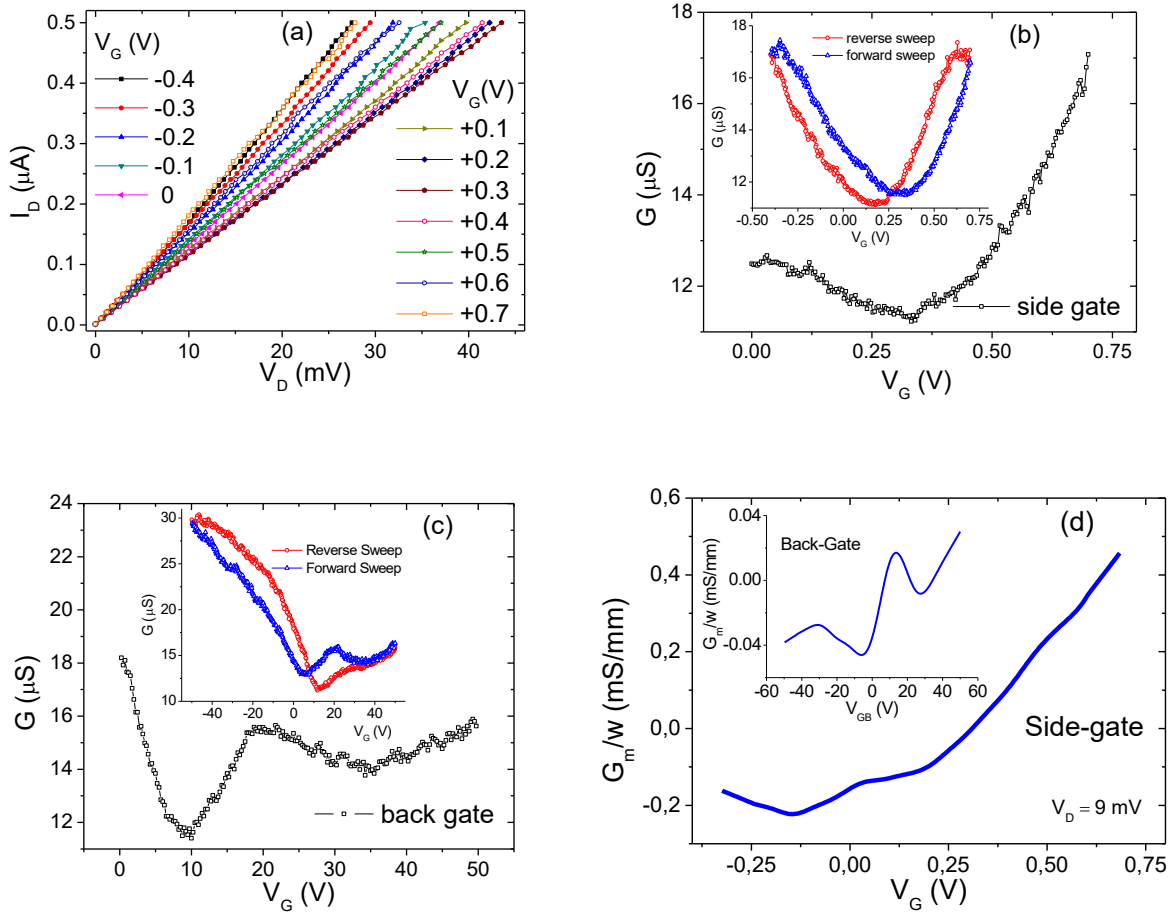

FIG. 3. Electrical characteristics of transistor T1 of Fig. 2. (a) Output characteristics ($I_D$-$V_D$ curves) for floating back-gate. (b) Transfer characteristic (G-$V_G$ curve) obtained at $V_D$ =9 mV and floating back-gate. (c) Transfer characteristic as a function of the back-gate for floating side-gate. Insets of

(b) and (c) show the transfer characteristics in a reverse-forward loop. (d) Side-gate and back gate trans-conductance, normalized by channel width.

To gain insight on the side-gate voltage that the device can withstand and to investigate the dielectric rigidity of the $SiO_2$/vacuum gate dielectric, we measured the planar current across the channel-gate gap till the appearance of a breakdown and beyond it. Such experiment was realized in high vacuum to remove surface moisture and adsorbates, which could provide extra leakage paths.

We started checking the vertical leakage, which is the current between the channel or the side-gate and the Si substrate, and we found it below the noise limit of 100 pA of our setup for biases up to 50 V. We did not perform measurements at higher voltages to avoid long and aggressive electrical stresses, which could trigger $SiO_2$ degradation mechanisms. For the same reason, to apply a higher voltage in the planar direction, between the channel and the side-gate, we decided to bias the back-gate at 50 V and then ramp the drain voltage $V_D$ up to 100 V, while the side-gate was grounded (Fig. 4(a)). In such a way, the maximum vertical stress was never higher than 50 V; besides, the back gate bias made graphene n-type, a favorable condition for leakage and electron emission measurements.

Figure 4(b) shows the current in the planar direction measured across the 100 nm gap between side-gate and channel. The current is below the noise floor of the experimental setup up to ~15 V and increases rapidly at higher $V_D$, while the back gate current keeps practically constant below 100 pA. The curve labelled as 2$^{nd}$ sweep in Fig. 4(b) represents the first full sweep after a few stabilizing cycles at lower voltages. The current emerges from the noise floor and increases over 3 decades up to ~60 V, before going through a dramatic change with a steeper rise up to 10 µA at 70 V. After that, a slow degradation happens. The appearance of a degradation mechanism is confirmed by the following sweep (3$^{rd}$ sweep), where the current appears at a higher voltage, $V_D$ ~35 V, and increases steadily and at a similar rate of the first part of the 2$^{nd}$ cycle without any significant change.

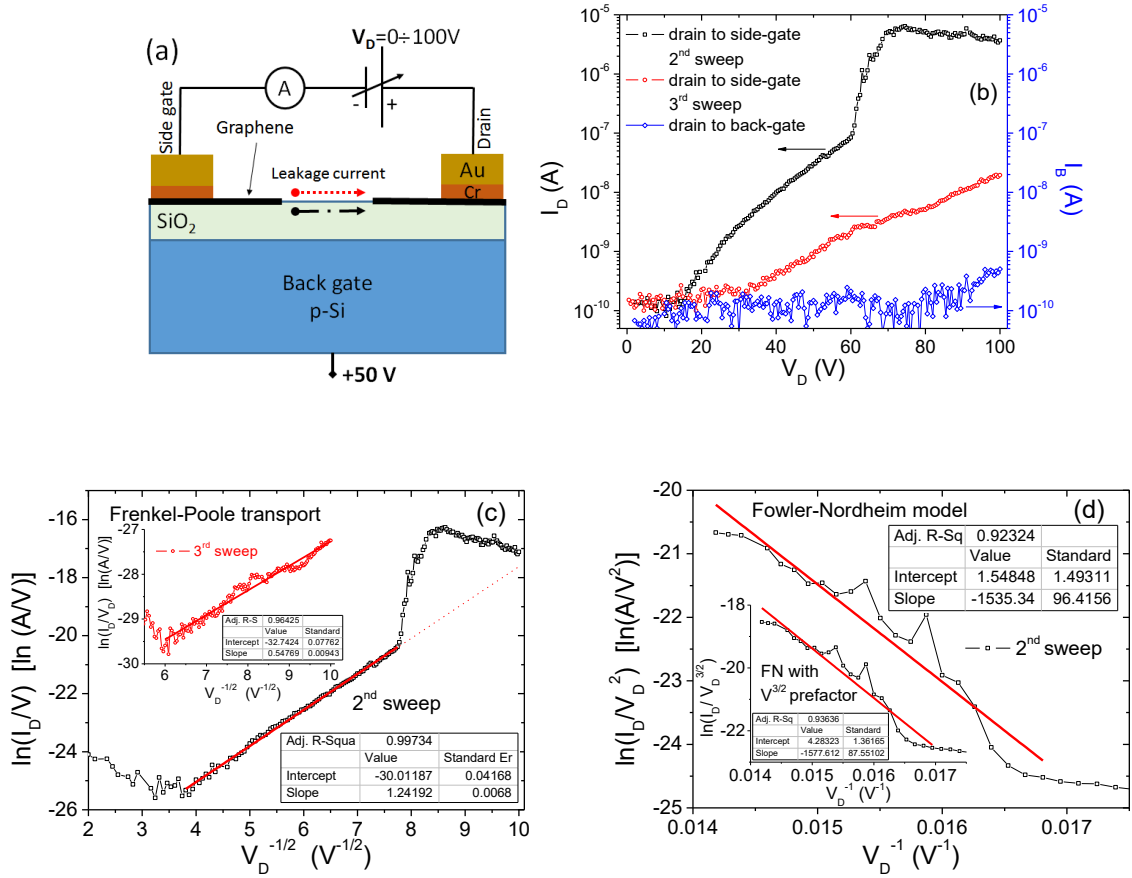

FIG. 4. (a) Measurement schematic of the side gate-channel leakage current. Measurements were performed in a SEM chamber at high vacuum ($10^{-6}$ Torr). (b) Planar current between channel and side-gate ($I_D$, left axis) and back-gate current ($I_B$, right axis) vs. drain bias ($V_D$) for the transistor T1 of Fig. 2(b). The 2$^{nd}$ sweep shows two transport regimes, before a high current degradation occurring at $V_D > 70$ V and resulting in reduced current in the following 3$^{rd}$ sweep. (c) Fit of Frenkel-Poole model to data of Fig. (b). The model is in excellent agreement with experimental data in the range 15-60V for 2$^{nd}$ sweep and in the range 35-100 V for 3$^{rd}$ sweep. (d) Fit of Fowler-Nordheim field emission model to the data of the 2$^{nd}$ sweep in the range 60-70 V.

The rapid increase of the current is typical of tunneling phenomena. However, the slower growing rate of the current below 60 V in the 2$^{nd}$ sweep, and after degradation in the 3$^{rd}$ cycle, seems to indicate that a different mechanism is taking place.

We compared the experimental data of figure 4(b) to the predictions of different transport models, as thermionic, Fowler-Nordheim and Frenkel-Poole emission or ohmic and space-charge-limited conduction [39-40]. Differently from Wang et al. [19], who reported current-voltage characteristics governed by the space-charge limited flow of current at low biases, on a device with similar layout but more complex fabrication process, we found that a far-better fit for $V_D<60$ V is provided by the Frenkel-Poole model:

$$I_D \propto V_D \exp\left(\frac{q}{kT}\left(2a\sqrt{V_D} - \phi_B\right)\right) \quad (1)$$

where q is the electron charge, k is the Boltzmann constant, $T$ the temperature, $a = \sqrt{q/(4\pi\varepsilon_0\varepsilon_r d)}$ is the Frenkel-Poole constant, $\varepsilon_r$ the dielectric constant of $SiO_2$ and $\phi_B$ is the trap barrier. The good fit is shown in the plot of $\ln(I_D/V_D)$ vs. $\sqrt{V_D}$ of Fig. 4(c), corresponding to a straight-line whose intercept can be used to evaluate $\phi_B \sim 0.8$ V, both when using the 2$^{nd}$ or the 3$^{rd}$ sweep. According to Frenkel-Poole model, electrons can be injected in trap states in the bandgap of the $SiO_2$, where they can move in a sequence of trapping and de-trapping events, facilitated by the electric field, which reduces the barrier on one side of the trap. For the Frenkel-Poole (FP) conduction mechanism to occur the trap must be neutral when filled with an electron, and positively charged when the electron is emitted, the interaction between positively charged trap and electron giving rise to a Coulombic barrier.

At higher voltages, $V_D>60$ V, the Fowler-Nordheim (FN) emission in vacuum becomes the model giving the best description of the data. The FN current is described by [20]

$$I \propto a\frac{1}{\phi}\left(\frac{\beta V}{d}\right)^2 \exp\left(-\frac{b\phi^{3/2}d}{\beta V}\right) \quad (2)$$

where $a = 1.54 \times 10^{-6} A \cdot eV \cdot V^{-2}$ and $b = 6.83 \times 10^{9} eV^{3/2} \cdot V/m$ are constants, $\phi$ is the barrier height (graphene workfunction) and $\beta$ is the field enhancement factor on the sharp edge of graphene. According to eq. (2), a plot of $\ln(I_D/V_D)$ vs $1/V_D$ (Fig. 4(d)) is straight-line whose slope and intercept are related to $\beta$ and $\phi$. Even though eq. (2) is typically used, it has been suggested that the pre-factor $V^2$ should be replaced by $V^{3/2}$ for graphene [41]. We checked FN model against other possible mechanisms and, despite the low statistics, we found that eq. (2), both in the original or modified version, is the closest to the experimental behavior. Hence, we concluded that FN injection in vacuum is the main leakage mechanism at high fields. The field enhancement factor at the edge of graphene, obtained from eq. (2) with $\phi$=4.5 eV, is ~4. Despite the atomically sharp edge only a modest amplification factor is achieved in our configuration, as confirmed also by a finite-element simulation of the field (MAXWELL software), shown in Fig. 5(a). The low value of $\beta$ is caused by the non-favorable edge-to-edge configuration of the two graphene flakes on the substrate and may be contributed by the small channel-to-gate distance, since the field enhancement factor is known to grow with the spacing between anode and the cathode [42-43].

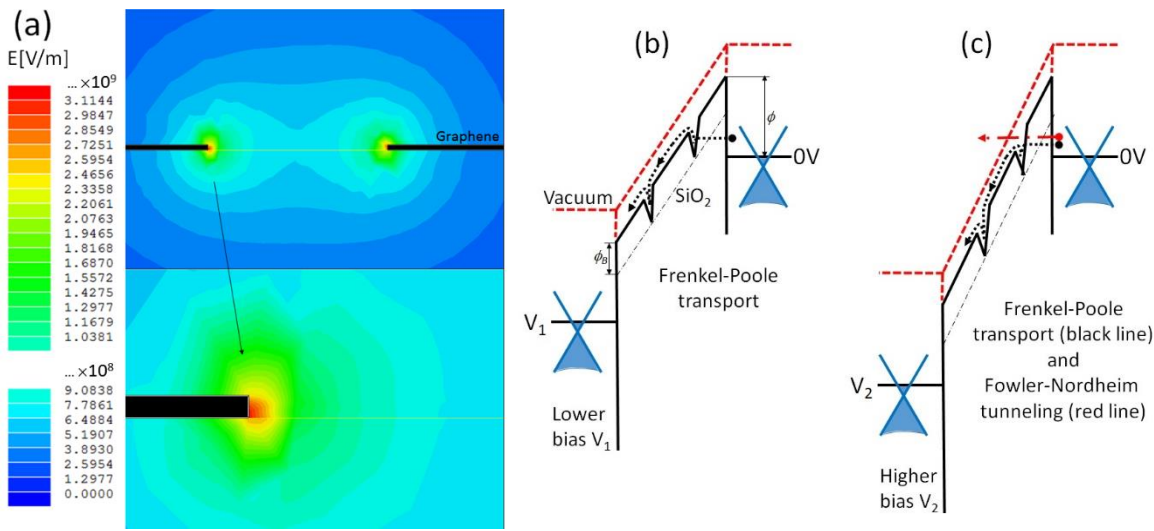

FIG. 5. (a) Electric field between the graphene flakes forming channel and side-gate. A modest amplification factor ($\beta$~4) is achieved at the graphene edge (the thickness of the graphene layers is set to 3 nm by software limitation). (b) Fowler-Nordheim tunneling and Frenkel-Poole transport at lower bias (< 60 V). (c) Fowler-Nordheim tunneling and transport in vacuum, which dominates over Frenkel-Poole transport, at higher bias.

As shown in Fig. 5, the electric field between the gate and channel lowers the graphene/SiO$_2$ and the graphene/vacuum barrier and enables two parallel paths for current flow (Fig. 4(a)). As sketched in the band diagrams of Fig. 5(b), at lower biases, electrons can be injected in local trap energy levels of the SiO$_2$ forbidden bandgap, and move through SiO$_2$ in a sequence of trapping and detrapping events. Traps are due to structural defects or stored charges. At higher biases, the additional Fowler-Nordheim (FN) tunneling in vacuum can take place (Fig. 5(c)) and become the dominating leakage mechanism, since electrons travel in vacuum from a graphene layer to the other without the capture and emission or any other scattering process limiting the current flow in the quasi-conduction band of SiO$_2$.

The degradation mechanism observed at high current (corresponding to $V_D$>70V) is likely caused by a modification of graphene edges, where joule heating can cause the sublimation of carbon atoms, similarly to what has been reported for carbon nanotubes [43-44] and for graphene at high bias [45]. This modification results in enhanced spacing and suppression of the field emission, which makes the leakage current to return to the FP regime.

## 4. Conclusions

In summary, we have fabricated side-gated all-graphene field effect transistors with gate-to-channel distance of 100 nm and channel width of 500 nm on SiO$_2$/Si substrates. We have shown that the side-gate is far more efficient than the back gate in modulating the channel conductance, with a 35% conductance swing over 0.5 V. We have studied the current leakage along the SiO$_2$/vacuum gap between the channel and the side-gate, and found that a rapidly increasing current appears for V > 15 V. We have clarified that the leakage current is caused by Frenkel-Poole transport at the SiO$_2$ surface at lower biases and becomes dominated by electron field emission in vacuum at higher bias. This study clarifies aspects of the side-gate approach, which is recently becoming popular for all-graphene nanoribbon transistors with high on-off ratio. It further provides background for the development of easy-to-fabricate planar field-emission devices for vacuum nanoelectronics.